\documentstyle[prd,aps,epsfig,preprint]{revtex}
\tighten
\begin{document}
\thispagestyle{empty}
\begin{center}
{\large\bf Transition Radiation of the Neutrino Toroid Dipole Moment}
\end{center}
\begin{center}
{Elena~N.~Bukina$^a$, Vladimir~M.~Dubovik$^a$ and
 Valentin~E.~Kuznetsov$^{a,b}$}
\hbox{\it $^a$Joint Institute for Nuclear Research,
          141980 Dubna, Moscow region, Russia}
\hbox{\it $^b$CERN, CH-1211, Geneva 23, Switzerland}
\end{center}
\date{\today}
\begin{abstract}
We discuss the transition radiation of a neutrino induced by its toroid
dipole moment ($\tau$) when crossing the interface between two
different media with different refraction indices.
A neutrino of 1 MeV energy emits approximately $10^{-40}$ keV at
$\tau=e\sqrt{2}G_F/\pi^2$. This effect depends very slightly
on the neutrino mass and has a finite value in the massless limit.
\end{abstract}


\section{Introduction}

$\quad$ The electromagnetic properties of a spin-$1/2$ charged particle
are described by four independent dipole moments while the neutrino
properties by three moments: magnetic ($\mu$), electric ($d$) and 
toroid ($\tau$) dipole moments~\cite{Dubovik1974,Kobzarev}. In the
Standard Model (SM) they are induced by radiative corrections and have
the following theoretical predictions for the electron neutrino
\cite{Lee,preprintDK,DKHEP}
 \footnote{\rm The numerical value of $\tau_{\nu_e}$ corresponds to
               toroid dipole moment of Majorana neutrino and is
               obtained summing over all contributions in eq. (16) of
               \cite{DKHEP}.}
 \begin{eqnarray}
 \mu_{\nu_e}\!&   =    &\!
               \frac{3eG_F m_{\nu_e}}{8\sqrt{2}\pi^2}=
              3\times10^{-19}\left(
               \frac{m_{\nu_e}}{1\;{\rm eV}}
               \right)\mu_B,
               \nonumber \\
 d_{\nu_e}  \!&   =    &\! 0,\quad E(0)\propto\Delta m,
               \nonumber \\
\tau_{\nu_e}\!&\approx &\!
              e\frac{\sqrt{2}G_F}{\pi^2}
               = e\cdot6.5\times10^{-34}\;{\rm cm}^2
               = 8.5\times10^{-13} \mu_B\lambdabar_e,
 \label{base}
 \end{eqnarray}
where $G_F$, $E$, $\mu_B$ and $\lambdabar_e$ are the Fermi constant, 
the electric form factor of the neutrino, the Bohr magneton and the
Compton wavelength of the electron, respectively.
At the same time, the experimental bounds
on these moments, which can be extracted in diverse ways
\cite{Salati}, are really poor \cite{PDG}.

The magnetic and electric dipole moments of neutrinos
are well known, but the third electromagnetic characteristic of
a neutrino, the toroid dipole moment (TDM), 
is still under discussion in the literature,
see for example \cite{preprintDK,DKHEP,BDK} and the references
therein. We know that the TDM is the electromagnetic
characteristic which the Dirac and Majorana neutrinos posses in
both the massive and massless limits.
In the non-relativistic limit  the interaction energy
${\cal H}=-\mbox{\boldmath $\tau$}\cdot{\bf J}
         =-\tau\varphi^{\dag}\mbox{\boldmath $\sigma$}
               \varphi\left({\rm curl}\,{\bf B}
               -\dot{{\bf E}}\right)$
represents a T-invariant electromagnetic interaction of the
particle induced by its TDM which does not conserve P- and
C-parity individually. It is useful to remark also that in the
massless limit, the electromagnetic properties of Dirac neutrinos
are represented by the TDM and the neutrino charge radius which
coincide numerically \cite{Kayser}. According 
to \cite{preprintDK,DKHEP}, the
spatial size of the toroid dipole moment (TDM) is formed by the
mass of the weak intermidiate boson $M_W$ and does not depend on
the inert mass of the particle under consideration. So in the
physical hierarchy the TDM is closer to the electric charge than
to the magnetic moment.

The permitted forms of coupling for the electromagnetic current
$J^{\rm EM}_\mu$ are
 \begin{eqnarray}
  J^{\rm EM}_\mu(q)_{\rm Dirac}   &=& \left[
  \overline{u}_f({\bf p}')\Gamma_\mu(q) u_i({\bf p})\right],
  \nonumber \\
  J^{\rm EM}_\mu(q)_{\rm Majorana}&=& J^{\rm EM}_\mu(q)_{\rm Dirac}
 +\left[
  \overline{v}_i({\bf p})\Gamma_\mu(q) v_f({\bf p}')\right] 
  \label{cur2},
 \end{eqnarray}
where the matrix elements are taken between the Dirac or Majorana 
neutrino states with  different masses.
A Lorentz-covariant structure of the dressed vertex operator
$\Gamma_\mu(q)$ in the toroid parametrization 
\cite{Dubovik1974,BDK} is given by 
 \begin{equation}
                    \Gamma_\mu(q) = F(q^2)\gamma_\mu
                    +M(q^2)\sigma_{\mu\nu}q^\nu
                    +E(q^2)\sigma_{\mu\nu}q^\nu\gamma_5
                    +i{\cal T}(q^2)\epsilon_{\mu \nu \lambda \rho}
                     P_{\nu}q_{\lambda}\gamma_{\rho},
 \label{vertex}
 \end{equation}
where $F$, $M$, $E$ and ${\cal T}$ are the normal, anomalous magnetic,
electric and toroid dipole form factors respectively,
$P_{\nu}= p_{\nu}+p'_{\nu}$ and $\epsilon_{\mu \nu \lambda \rho}$
is the antisymmetric tensor. In the anapole parametrization 
\cite{Zeld} the vertex operator reads
 \begin{equation}
                    \Gamma_\mu(q) = F(q^2)\gamma_\mu
                    +M(q^2)\sigma_{\mu\nu}q^\nu
                    +E(q^2)\sigma_{\mu\nu}q^\nu\gamma_5
                    +A(q^2)[q^2\gamma_\mu-\widehat{q}q_\mu]\gamma_5
 \label{vertex2}
 \end{equation}
where $A(q^2)$ is the anapole form factor. Using the following 
identity
 \begin{eqnarray}
    \overline{u}_f({\bf p}')\Bigl\{ \Delta m\sigma_{\mu\nu}q^\nu
 &+&\left(q^2\gamma_\mu-\widehat{q}q_\mu\right)
    \nonumber \\
 &-&i\varepsilon_{\mu\nu\lambda\rho}P^\nu q^\lambda
                  \gamma^\rho\gamma_5
                  \Bigr\}\gamma_5 u_i({\bf p}) = 0,
  \label{II.3}\end{eqnarray}
we see that the TDM and anapole coincide
in the static limit when the initial and final masses of neutrinos
are equal to each other \cite{Dubovik1974,BDK}.

It is easy to check, using CPT invariance of $\Gamma_\mu$ and C-, P-
and T-properties of each contribution in 
eqs. (\ref{vertex}, \ref{vertex2}),
that for the Majorana current only the toroid dipole form factor 
survives \cite{Kobzarev} and the value of the toroid dipole moment 
of the Dirac neutrino is just half of the Majorana one.
For the above reasons we have not specified the nature of the 
neutrino and as TDM is a more simple (multipolar) characteristic than 
anapole,  which has the composite structure as it follows from 
(\ref{II.3}), we shall subsequently only use the term TDM. 
In addition, in the forthcoming calculations the numerical value of 
TDM from eq. (\ref{base}) will be used \cite{preprintDK,DKHEP}.

If the toroid dipole moment is observable, what
physical consequences does it lead to?
Among the several possibilities are the Vavilov-Cherenkov and
transition radiations (TR) of particles induced by their dipole 
moments. This problem for the Dirac
neutrino with non-zero magnetic moment was considered in
\cite{GrimusNeufeld1993,Sakuda1995}. In 1985, Ginzburg and
Tsytovich \cite{GinzTsyt}, using a classical approach, showed
that the macroscopic toroid dipole moment moving in a medium
generates Vavilov-Cherenkov and TR radiations as well.
Here we present the first
quantitative discussion of the transition radiation of a neutrino
having non-zero TDM in the framework of quantum theory along the
lines of \cite{Sakuda1995}.

\section{Calculation of transition radiation \\
intensity}

\begin{figure}
\epsfig{file=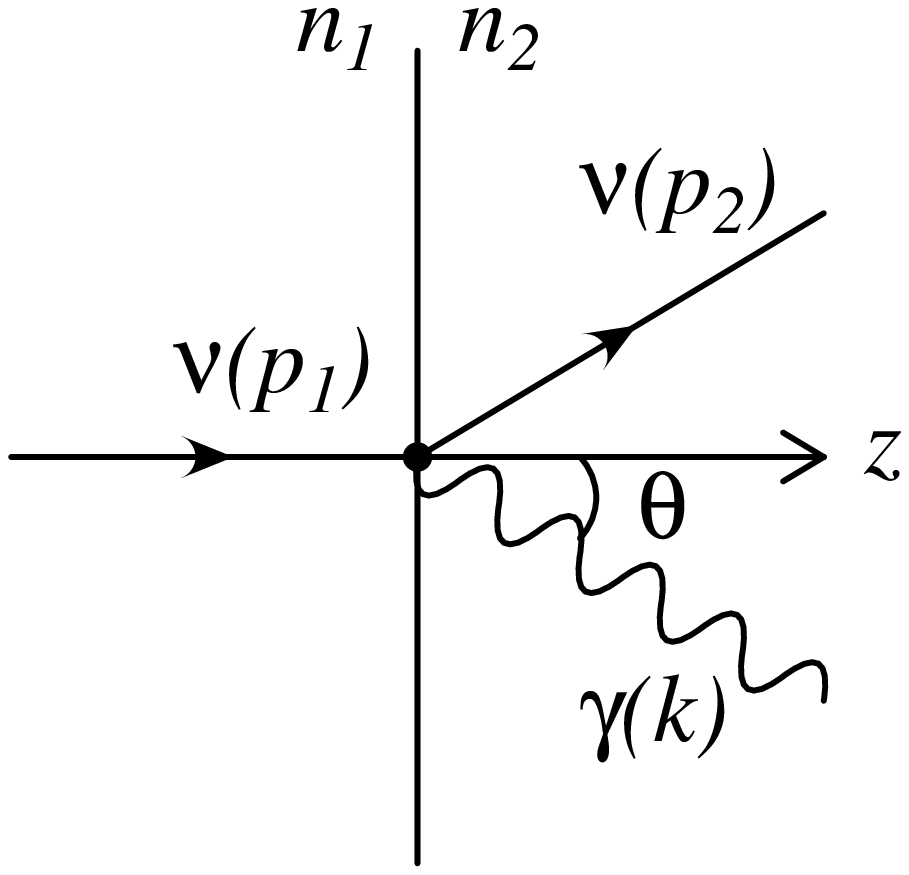,width=6cm}
\protect\caption{\label{Fig1}
        Transition radiation at the interface of two media:
        $\nu (p_1) \rightarrow \nu (p_{2}) + \gamma (k)$.
        The refractive index changes from $n_{1}$ to $n_{2}$ at $z=0$.}
\end{figure}

$\quad$ Let us consider a neutrino with non-zero toroid dipole moment
crossing the interface between two media, see Fig. \ref{Fig1},
with refraction indices $n_1$ and $n_2$ ($n_1\neq n_2$).
The electromagnetic interactions
of neutrinos is described by the Hamiltonian:
 \begin{eqnarray}
    {\cal H}_{\rm int} & = &
    i e{\cal T}(q^2)\overline{\psi}(x)\varepsilon_{\mu \nu \lambda
    \rho} P^{\nu} q^{\lambda} \gamma^{\rho} \psi(x){\cal A}^\mu(x),
                    \nonumber 
 \end{eqnarray}
using the identity
$ \varepsilon_{\mu \nu \lambda \rho} \gamma_{\rho} = \frac{i}{2}
\left( \gamma_{\mu} \gamma_{\nu} \gamma_{\lambda} -
\gamma_{\lambda} \gamma_{\nu} \gamma_{\mu} \right) \gamma_{5}$, we
%
we obtain
  \begin{eqnarray}
    {\cal H}_{\rm int} & \Rightarrow &
    e{\cal T}(q^2)\overline{\psi}(x)\gamma_\mu\gamma_5 \psi(x)
        \frac{\partial F^{\mu\nu}(x)}{\partial x^\nu}
                    \nonumber \\
                        & = &
    e{\cal T}(q^2)\overline{\psi}(x) \gamma_\mu
                                 \gamma_5\psi(x)J^\mu_{\rm ext}.
 \label{Eq1}
 \end{eqnarray}

Here $\psi$, ${\cal T}(q^2)$, $J^\mu_{\rm ext}$, ${\cal A}^\mu$
and $F^{\mu\nu}$ are the neutrino wave function, neutrino toroid
form factor, electromagnetic current, $4-$potential and tensor of
the electromagnetic field, respectively
(the Hamiltonian (\ref{Eq1}) was also obtained by Zel'dovich 
\cite{Zeld} using the anapole parametrization).

The transition $\nu(p_1)\rightarrow\nu(p_2)+\gamma(k)$ becomes
possible due to the TDM of the neutrino~\footnote{Recall that both 
the Vavilov-Cherenkov and transition radiations are not depend on 
the nature of a moving source~\cite{GinzTsyt}}.
In a medium with refraction index $n$,
the four-momentum vector of a photon is given by
$k^\mu=(\omega,{\bf k}),\quad |{\bf k}|=n\omega$ ($\omega$ is the
energy of a photon), and the transition probability reads
$\Gamma=|{\cal S}_{fi}|^2\frac{Vd^3{\bf p}_2}{(2\pi)^3}
        \frac{Vd^3{\bf k}}{(2\pi)^3}$, where
the transition matrix element is expressed as
\begin{eqnarray}
 |{\cal S}_{fi}|^2
           &   =  &(2\pi)^3\ell^2t\frac{m_\nu}{E_1V}\frac{m_\nu}{E_2V}
                          \frac{(1-n^2)^2\omega^4}{2\omega n^2 V}
                          \nonumber \\
           &\times&
                          \delta(p_{1x,y}-p_{2x,y}-k_{x,y})
                          \delta(E_1-E_2-\omega)
                          \nonumber \\
           &\times& \left|\int^{\ell/2}_{-\ell/2}dz
                          \exp[i(p_{1z}-p_{2z}-k_z)z]{\cal M}_{fi}
                    \right|^2.
\label{eq:ma}\end{eqnarray}
 Here
${\cal M}_{fi}=e{\cal T}(0)\overline{u}_2
                 \widehat{\varepsilon}\gamma_5 u_1$
is the amplitude, and $t$, $\ell$ and $V=\ell^3$ denote time,
length and volume of the transition region, respectively, and
$\ell=\beta t$, where $\beta=p/E$ is the velocity of the neutrino.
The phase of the integrand in (\ref{eq:ma}) defines
the formation-zone length of the medium as
$$Z(n)=(p_{1z}-p_{2z}-k_z)^{-1}=
       (p_{1z}-p_{2z}-n\omega\cos\theta)^{-1},$$
where $\theta$ is the angle between the photon and the direction of 
the incident neutrino. The details of further calculations are
the same as in~\cite{Sakuda1995}, and here we present only final
results for the energy intensity $S$ per interface
\begin{equation}
   \frac{d^2 S}{d\theta d\omega}=
   \frac{{\cal T}^2(0) \omega^6\sin\theta}{8\pi^2}(R_1^2-R_2^2)
   \left\{2\sin^2\theta\left(1+\frac{n\omega\cos\theta}{p_{2z}}\right)
        +\frac{E_\nu E_2}{pp_{2z}}-1+\frac{m_\nu^2}{pp_{2z}}\right\},
\label{dS}
\end{equation}
\begin{flushleft}
where
\end{flushleft}
\begin{equation}
   R_i=\frac{1-n_i^2}{n_i}\frac{1}{p-p_{2z}-n_i\omega\cos\theta},\quad
   S=\int_0^{E_\nu-m_\nu}d\omega\int_0^{\theta_{\rm max}}
     \frac{d^2 S}{d\theta d\omega},
\label{S}
\end{equation}
and
$$p_1^\mu=(E_\nu,0,0,p),\quad
  p_{2z}=\sqrt{E_2^2-m_\nu^2-n^2\omega^2\sin^2\theta},
         \quad E_2=E_\nu-\omega.$$

\begin{figure}\mbox{\epsfig{file=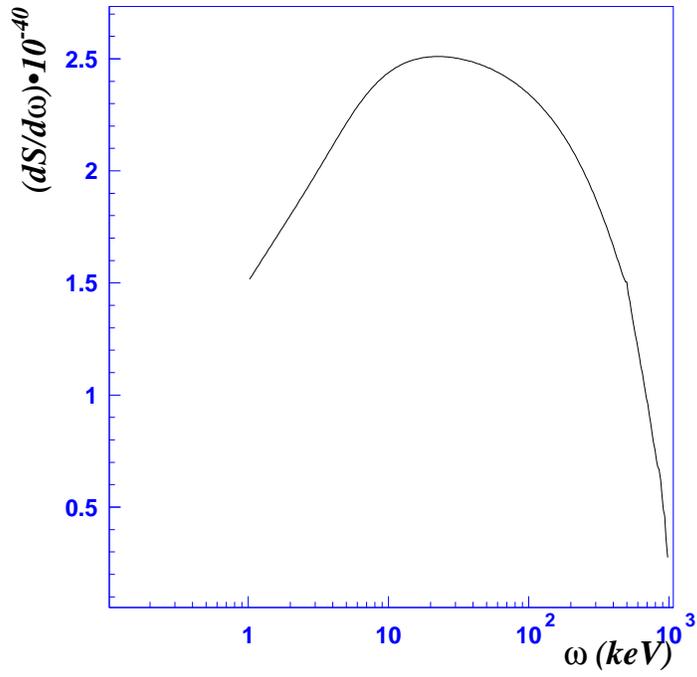,width=10cm}}
\protect\caption{\label{Fig2}
        Energy intensity distribution of the transition radiation
        of a toroid dipole moment of neutrino as a function of
        photon energy.}
\end{figure}

\begin{figure}\mbox{\epsfig{file=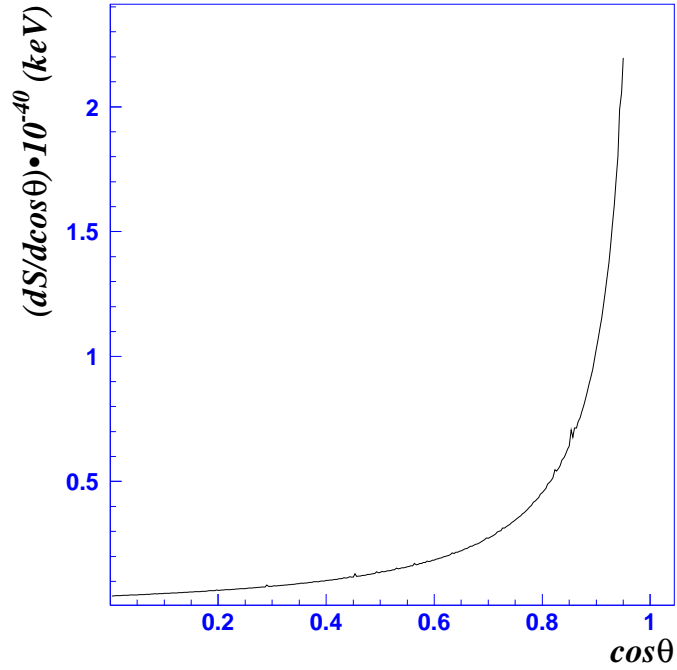,width=10cm}}
\protect\caption{\label{Fig3}
        Angular distribution of total TR energy as a function of
        $\cos\theta$.}
\end{figure}

\begin{figure}\mbox{\epsfig{file=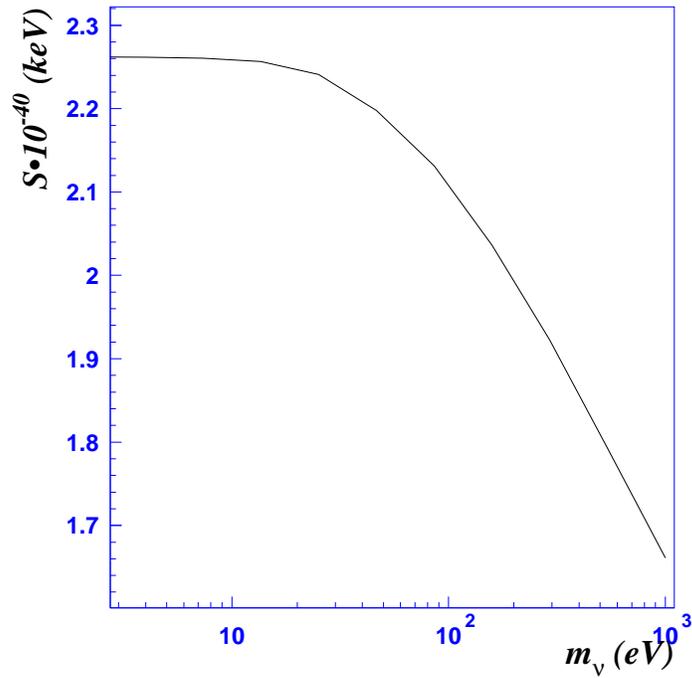,width=10cm}}
\protect\caption{\label{Fig4}
        Total energy in medium-vacuum transition, $n_2=0$ and
        $n_i(\omega)\simeq 1-\omega_i^2/2\omega^2$ for
        $\omega\gg\omega_i$ and $E_\nu=1$ MeV.}
\end{figure}

Using the numerical value (\ref{base}), $\tau_{\nu_e}=e{\cal T}(0)$ 
with ${\cal T}(0)=\sqrt{2}G_F/\pi^2$, and assuming
that the refractive index can be expressed as
$n_i(\omega)=1-\omega_i^2/2\omega^2$ for $\omega\gg \omega_i$
($\omega_i$ is the plasma frequency) for a medium-vacuum
transition ($\omega_2=0$, $R_2=0$), we present the energy spectrum
and angular distribution in Figs. \ref{Fig2} and \ref{Fig3}.
The total energy loss of the neutrino has been computed numerically
and is shown as a function of the neutrino mass for $E_\nu=1$ MeV
in Fig. \ref{Fig4}. For $m_\nu<10$ eV
the TR energy is approximately constant and equals
$S\simeq 2\times10^{-40}$ keV. Because of the finite value of the TDM
for massless neutrinos \cite{preprintDK,DKHEP}, the TR does not
vanish in this limit and has the value
$S\Bigr|_{m_\nu=0}=2.26\times10^{-40}$ keV.

In order to estimate the magnitude of this effect, let us consider
a transition radiation detector (TRD) which can be used
to measure experimentally such transition radiation of neutrinos.
The TRD consists of sets ($N_1$) of ``radiator'' and xenon-gas
chambers, where one radiator typically comprises of a few hundred
layers ($N_2$) of a polypropylene film ($\omega_p=20$ eV) and a
gas ($\omega_p\ll 1$ eV). The total energy deposition ($W$) in the TRD
is given by
\[ W=S\cdot F_\nu\cdot A^2\cdot T\cdot N_1\cdot N_2, \]
where $F_\nu$ is the neutrino flux, $A$ is the TRD area and $T$ is the
time of measurement. For example, we have taken the neutrino flux
coming from a nuclear reactor,
$F_\nu\sim10^{13}\overline{\nu}_e/{\rm cm}^2\;{\rm sec}$,
with the energy of $E_\nu=1$ MeV and TRD parameters as: $A=10$ m$^2$
$N_1=10$ sets and $N_2=10^4$ layers. The total energy deposition for
the TDM $\tau_{\nu_e}=e\sqrt{2}G_F/\pi^2$ with TR energy $S=10^{-40}$
keV is
$$
  W=3\times10^{-10}\left(\frac{T}{1\; {\rm year}}\right).
$$
Unfortunately, this value is extremely small and cannot be
extracted from the background of an experimental setup.
But this small radiation which always exist for both massive
and massless neutrinos, may have interesting consequencies
in astrophysics.

\section{Conclusion}

$\quad$ In summary, we have calculated
the transition radiation of a neutrino induced by its toroid dipole
moment in the framework of quantum theory. Since the TDM of a neutrino 
is nonzero in the massless limit \cite{preprintDK,DKHEP}, the 
corresponding TR energy is also nonzero and equals 
$S\simeq 2.26\times10^{-40}$ keV for
$\tau_{\nu_e}=e\sqrt{2}G_F/\pi^2$.~\footnote{The transition radiation
induced by the neutrino magnetic moment disappeares in the massless 
limit since $\mu_\nu\sim m_\nu$.} 
In addition the TDM is the weak-electromagnetic characteristic which 
both Dirac and Majorana neutrinos posses, therefore the transition 
radiation induced by TDM exist indepedently on the nature of the 
neutrino and its mass.

It is highly plausible that we will stand
face-to-face with the dramatic circumstances if
either the Dirac neutrinos possess negligible masses and their
magnetic moments are also small, or if all neutrinos have a Majorana 
nature. Then the unique electromagnetic characteristic of such 
neutrinos will be the toroid moment and we will be forced to seek for 
some exotic effects generated by it.
For instance, if the neutrino is a massless particle then measurement
of the transition radiation can be used as a tool to distinguish the
nature of the neutrino (since the Dirac TDM is half of the Majorana one
and as the energy intensity is proportional to the square of the TDM
(\ref{dS}), the TR of Dirac neutrino is 1/4 of the Majorana one).

It is interesting to note that TR energy of the order of
$10^{-40}$ keV for a neutrino with TDM  corresponds to the TR energy
of a neutrino with anomalous magnetic moment
$\mu_\nu\sim10^{-15}\mu_B$ for $m_\nu=1$ eV \cite{Sakuda1995}.
Such TR of neutrinos induced by their TDMs may have interesting
implications for astrophysics as well as the early Universe. However,
the conclusions about the magnitude of these effects requires
further investigation.

\newpage
\thispagestyle{empty}


\newpage
\thispagestyle{empty}


\end{document}